\documentclass[,final]{aipproc}
\layoutstyle{6x9}

\begin{document}

\title{Transport properties in bilayer Quantum Hall systems in the presence of a topological defect}


\author{Gerardo Cristofano}{
  address={Dipartimento di Scienze Fisiche, Università di Napoli "Federico II" and INFN, Sezione di Napoli, Complesso Universitario di Monte S. Angelo, Via Cinthia, I-80126 Napoli, Italy}
}

\author{Vincenzo Marotta}{
  address={Dipartimento di Scienze Fisiche, Università di Napoli "Federico II" and INFN, Sezione di Napoli, Complesso Universitario di Monte S. Angelo, Via Cinthia, I-80126 Napoli, Italy}
}

\author{Adele Naddeo}{
  address={Dipartimento di Fisica "E. R. Caianiello", Università di Salerno and CNISM, Unità di Ricerca di Salerno, I-84081 Baronissi (SA), Italy}
}

\author{Giuliano Niccoli}{
  address={Sissa and INFN, Sezione di Trieste - Via Beirut 1 - 34100 Trieste, Italy}
}
\begin{abstract}
Following a suggestion given in \cite{noi}, we show how a bilayer
Quantum Hall system at fillings $\nu =\frac{1}{p+1}$ can exhibit a
point-like topological defect in its edge state structure. Indeed
our CFT theory for such a system, the Twisted Model (TM), gives
rise in a natural way to such a feature in the twisted sector. Our
results are in agreement with recent experimental findings
\cite{deviatov1} which evidence the presence of a topological
defect in the transport properties of the bilayer system.
\end{abstract}
\maketitle

\section{Introduction}

Recently bilayer quantum Hall systems have been widely
investigated theoretically as well as experimentally \cite{teoria,
esperim}. Indeed, when tunneling between the layers is weak, the
quantum Hall bilayer state can be viewed as arising from the
condensation of an excitonic superfluid in which an electron in
one layer is paired with a hole in the other layer. The
uncertainty principle makes it impossible to tell which layer
either component of this composite boson is in. Equivalently the
system may be regarded as a ferromagnet in which all electrons
appear in a coherent superposition of the \textquotedblright
pseudospin\textquotedblright\ eigenstates which encode the layer
degree of freedom \cite{YangMoon}\cite {girvin1}. The phase
variable of such a superposition fixes the orientation of the
pseudospin magnetic moment and its spatial variations govern the
low energy excitations in the system. Since Halperin work
\cite{Halperin} the concept of edge states was introduced in order
to describe transport phenomena in two dimensional electron
systems. They arise in a quantized magnetic field at the
intersections of the Fermi level with different Landau levels,
which are bent up by the edge potential. In particular the
formation of a topological defect has been predicted to occur when
two edge states with different spins locally switch their
positions and thus cross each other at two or more points
\cite{edge1}. More interesting features take place in the
transport properties of bilayer systems when also pseudospin
(related to the layer index) is involved
\cite{girvin1}\cite{sarma}. Recently the presence of edge state
crossings and thus of topological defects has been experimentally
evidenced in such systems in a quasi-Corbino geometry
\cite{deviatov2} at filling $\nu =3$ \cite{deviatov1} by means of
a selective population technique. In particular the application of
a suitable gate voltage $V_{g}$ and of a magnetic field drives the
bilayer in different pseudospin states in the gated and ungated
regions, so producing a crossing of the edge states which has been
detected in the transport properties. The net result is a linear
$I-V$ characteristics for the electric transport between two
different edges. Because the gate-gap width is smaller than the
characteristic equilibration lengths in such a transport between
the edge states, it has been argued that a defect must be present,
which couples different edge states but only with the same spin in
the gate-gap. Such a picture can be destroyed by an in-plane
magnetic field component which washes out the above crossing; the
$I-V$ curves become then strongly non-linear so signaling the
merging of a tunneling process. All the above features in the
$I-V$ characteristics appear to be the fingerprints of the
presence of a topological defect induced by the different
pseudospin configurations in bilayer quantum Hall systems
\cite{deviatov1}.

In this contribution we address theoretically the issue of the
presence of topological defects in the Conformal Field Theory
(CFT) description of the edge states of bilayer quantum Hall
systems in a wide class of filling factors, and in particular the
paired states ones, in the framework of our TM approach
\cite{noi}. In particular we show how such a feature arises in a
very natural way in the twisted sector of our theory, as a result of the $m$%
-reduction technique \cite{cgm1}\cite{cgm4}. The transport
properties of bilayer systems will be investigated by studying the
properties under magnetic translations of the characters of the
different sectors, which describe its different non perturbative
ground states. The paper is organized as follows. In Section 2 we
recall those aspects of our $m$-reduction procedure which turn out
to be relevant for the description of bilayer systems with
topological defects. In Section 3 we study the transport
properties of such systems by means of magnetic translations
pointing out how they arise from Laughlin gauge argument. Finally
some conclusions and outlooks are given.
\newline

\section{$m$-reduction technique: a description of bilayer systems with topological defects}

The $m$-reduction technique is based on the simple observation
that for any CFT (mother) exists a class of sub-theories
parameterized by an integer $m$ with the same symmetry but
different representations. The resulting theory
(daughter) has the same algebraic structure but a different central charge $%
c_{m}=mc$. To obtain the generators of the algebra in the new
theory we need to extract the modes which are multiple of the
integer $m$. These can be used to reconstruct the primary fields
of the daughter CFT. This technique can be generalized and applied
to any extended chiral algebra which includes the Virasoro one.
Indeed the $m$-reduction preserves the commutation relations
between the algebra generators but modifies the central extension
(i.e. the level for the WZW models). In particular this implies
that the number of primary fields gets modified. Its application
to the QHE\ arises by the incompressibility of the Hall fluid
droplet at the plateau, which implies its invariance under the
$W_{1+\infty }$ algebra at the different fillings, and by the
property of the $m$-reduction procedure to obtain a daughter CFT
with the same\ $W_{1+\infty }$ invariance property of the mother
theory. Thus the $m$-reduction furnishes automatically a mapping
between different incompressible plateaux of the QHF.

The general characteristics of the daughter theory is the presence
of twisted boundary conditions (TBC) which are induced on the
component fields. It is illuminating to give a geometric
interpretation of that in terms of the covering on a $m$-sheeted
surface or complex curve with branch-cuts, see Fig. 1.

\begin{figure}
  \includegraphics[height=.3\textheight]{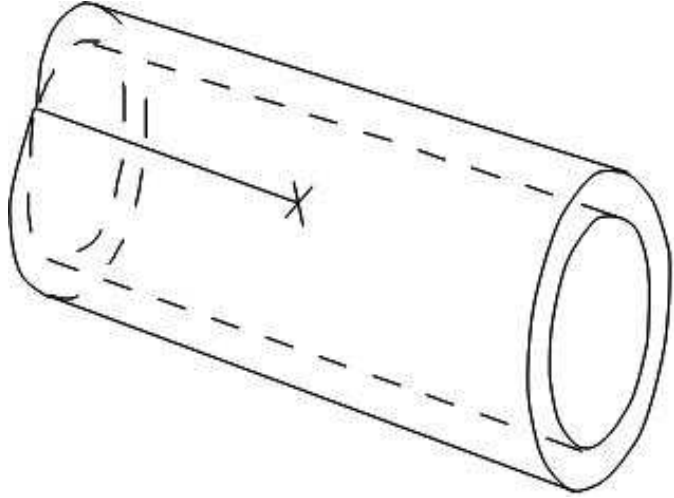}
  \caption{The boundaries of the 2-covered cylinder can be viewed as
different configurations of the QHF edges described by the
2-reduced CFT.}
\end{figure}

Indeed the fields which are defined on the left boundary have TBC
while the fields defined on the right one have periodic boundary
conditions (PBC). We point out that fields with TBC describe
elegantly the crossing between the layers as a consequence of the
presence of a branch-cut. We find different sectors on the torus
corresponding to different boundary conditions on the cylinder.
Finally we recognize the daughter theory as an orbifold of the
usual CFT describing the QHF at a given plateau. The two sheets
simulate the two-layers system and the branch cut represents TBC
which emerge from the interaction with a localized defect on the
edge. This is a key feature of our construction, as we will point
out in the following.

In order to see how the $m$-reduction procedure works on the plane \cite%
{cgm1} and on the torus \cite{cgm4} and how it gives rise to the
edge state coupling via a topological defect, let us focus on the
paired states fillings in the special $m=2$ case since we are
interested in a system consisting of two parallel layers of $2D$
electrons gas in a strong perpendicular magnetic field. The
filling factor $\nu ^{(a)}=\frac{1}{2p+2}$ is the same for the two
$a=1$, $2$ layers while the total filling is $\nu =\nu ^{(1)}+\nu
^{(2)}=\frac{1}{p+1}$. We point out that our results can be
generalized to any bilayer system. The simplest abelian quantum
Hall state in the disc topology is written as a generalization of
the analytic part of the Laughlin wave function \cite{Halperin}:
\begin{equation}
f\left( z_{i}^{(a)}\right) =\prod_{a=1,2}\prod_{i<j}\left(
z_{i}^{(a)}-z_{j}^{(a)}\right) ^{2+p}\prod_{i,j}\left(
z_{i}^{(1)}-z_{j}^{(2)}\right) ^{p} ;  \label{halp}
\end{equation}%
in particular, for $p=0$ it describes the bosonic $220$ state and, for $p=1$%
, the fermionic $331$ one. The CFT description for such a system
can be given in terms of two compactified chiral bosons $Q^{(a)}$
with central charge $c=2$. A similar result can be obtained for
filling $\nu ^{\left( a\right) }=1/(2p+1)$ (Jain series).

In order to construct the fields $Q^{(a)}$ for the TM, let us
start from the
bosonic \textquotedblleft Laughlin\textquotedblright\ filling $\nu =1/2(p+1)$%
, described by a CFT with $c=1$ in terms of a scalar chiral field
$Q$
compactified on a circle with radius $R^{2}=1/\nu =2(p+1)$ (or its dual $%
R^{2}=2/(p+1)$). It is explicitly given by:
\begin{equation}
Q(z)=q-i\,p\,lnz+i\sum_{n\neq 0}\frac{a_{n}}{n}z^{-n}
\label{modes}
\end{equation}%
with $a_{n}$, $q$ and $p$ satisfying the commutation relations
$\left[
a_{n},a_{n^{\prime }}\right] =n\delta _{n,n^{\prime }}$ and $\left[ q,p%
\right] =i$. From such a CFT (mother theory), using the
$m$-reduction procedure, which consists in considering the
subalgebra generated only by the modes in eq. (\ref{modes}) which
are a multiple of the integer $m$, we get a $c=2$ orbifold CFT
(daughter theory, i.e. the TM) which describes the LLL dynamics.
Then the fields in the mother CFT can be organized into components
which have well defined transformation properties under the
discrete $Z_{2}$ (twist) group, which is a symmetry of the TM. Its
primary fields content can be expressed in terms of a
$Z_{2}$-invariant scalar field $X(z)$, given by
\begin{equation}
X(z)=\frac{1}{2}\left( Q^{(1)}(z)+Q^{(2)}(-z)\right) ,  \label{X}
\end{equation}
describing the electrically charged sector of the new filling, and
a twisted field
\begin{equation}
\phi (z)=\frac{1}{2}\left( Q^{(1)}(z)-Q^{(2)}(-z)\right) ,
\label{phi}
\end{equation}
which satisfies the twisted boundary conditions $\phi (e^{i\pi
}z)=-\phi (z)$ and describes the neutral sector \cite{cgm1}. Such
TBC signal the presence
of a topological defect which couples, in general, the $m$ edges in a $m$%
-layers system. In the bilayer system ($m=2$) we get a crossing
between the two edges as sketched in Fig. 2.

\begin{figure}
  \includegraphics[height=.2\textheight]{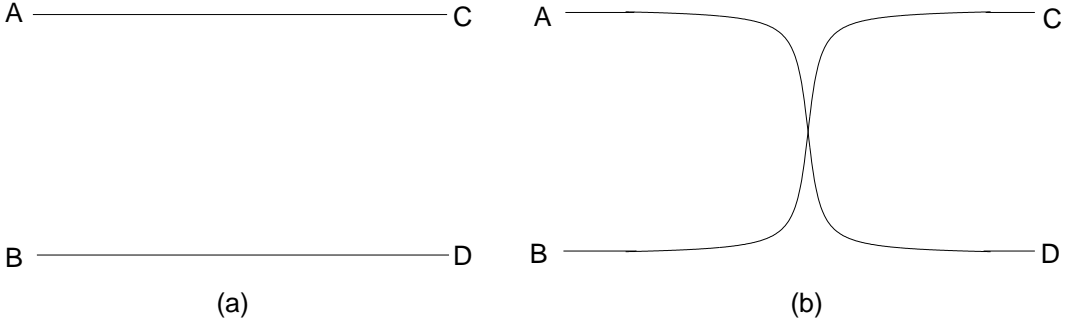}
  \caption{The
bilayer system, (a) without the topological defect (PBC), (b) with
the topological defect (TBC).}
\end{figure}

The chiral fields $Q^{(a)}$, defined on a single layer $a=1$, $2$,
due to the boundary conditions imposed upon them by the orbifold
construction, can be thought of as components of a unique
\textquotedblleft
boson\textquotedblright\ defined on a double covering of the disc (layer) ($%
z_{i}^{(1)}=-z_{i}^{(2)}=z_{i}$). As a consequence of such a
construction
the two layers system becomes equivalent to one-layer QHF and the $X$ and $%
\phi $ fields defined in eqs. (\ref{X}) and (\ref{phi})
diagonalize the interlayer interaction. In particular the $X$
field carries the total charge with velocity $v_{X},$ while $\phi
$ carries the charge difference of the two edges with velocity
$v_{\phi }$ i.e. no charge, being the number of electrons the same
for each layer (balanced system).

The TM primary fields are composite operators and, on the torus,
they are described in terms of the conformal blocks (or
characters). Furthermore a topological defect appears in our
formalism, being induced by the different isospin configurations
on the two layers, which naturally result from our $m$-reduction
procedure. The effect of a topological defect in a
quantum Hall fluid has been recently evidenced in experimental findings \cite%
{deviatov1}, as we will show in the following. In the presence of
a localized defect two phenomena can take place. A tunneling of
edge quasi-particles at point $x_{0}$, described by\ a boundary
term Hamiltonian such as:
\begin{equation}
H_{P}=-t_{P}\cos \left( Q^{(1)}-Q^{(2)}\right) \delta \left(
x_{0}\right) . \label{tunnel}
\end{equation}%
A second mechanism producing a current flow between the two edges
can be addressed to a localized crossing of the edges,\ which can
be represented by a boundary term:
\begin{equation}
H_{\beta }=\beta \left( Q^{(1)}\partial
_{t}Q^{(2)}-Q^{(2)}\partial _{t}Q^{(1)}\right) \delta \left(
x_{0}\right) ,  \label{magterm}
\end{equation}%
where $\beta =0(1/2)$ for PBC (TBC) respectively (see Fig.2). The
full Hamiltonian can be written as:
\begin{eqnarray}
H &=&\frac{1}{2}\sum_{a=1,2}\left[ \left( \Pi ^{(a)}\right)
^{2}+\left(
\partial _{x}Q^{(a)}\right) ^{2}\right] +H_{P}+H_{\beta }  \nonumber \\
&+&eV\partial _{t}\left( Q^{(1)}-Q^{(2)}\right) ,  \label{fullham}
\end{eqnarray}%
where $\Pi ^{(a)}$ is the momentum conjugate to $Q^{(a)}.$ We
recognize a kinetic term for the two bosonic fields
$Q^{(a)},a=1,2$, a boundary
tunneling term which implements the locally applied gate voltage $%
V_{g}=t_{P}\delta \left( x_{0}\right) $, a boundary magnetic term \cite%
{maldacena} which couples the two fields introducing a topological
defect (see ref. \cite{noi} for details) and a voltage switching
term between the two layers. The last term contains an irrelevant
operator, so it doesn't change the central charge: it behaves as a
boundary condition changing operator allowing for the flow from a
boundary state to another one.
Introducing the charged and neutral fields $X$ and $\phi $ defined in eqs. (%
\ref{X}) and (\ref{phi}) we clearly see that the last term in the
Hamiltonian is proportional to the neutral current, so it
contributes to unbalance the system. Therefore edge-crossing can
be described by TBC on the $\phi $ field induced by the boundary
magnetic term of eq. (\ref{magterm}).

\section{Study of transport properties: magnetic translations and Laughlin gauge argument}

The transport properties of the bilayer system under study can be
investigated by the application of different chemical potentials
between the terminals of Fig. 2, that we represent by the matrix
$V=\left(
\begin{array}{cc}
V_{AC} & V_{AD} \\
V_{BC} & V_{BD}%
\end{array}%
\right) $ with entries $V_{IJ}$, the potentials between the $I$
and $J$ terminals. Let us consider the following two cases, the
one in which the
transport of electrons is on the two independent edges through the points $%
A-C-A$, $B-D-B$ in the non crossed case (PBC see Fig. 2a) and the
one in which the transport is through the points $A-D-B-C-A$ in
the crossed edge case (TBC see Fig. 2b). In both cases there is no
tunneling ($t_{p}=0$) and they correspond respectively to the
diagonal (i.e. $V_{AD}=V_{BC}=0$) and to the anti-diagonal (i.e.
$V_{AC}=V_{BD}=0$) configurations respectively.

In a closed geometry, such as that of a torus, they can be induced
by adiabatic magnetic flux insertion through a cycle of the torus
(i.e. $A$ or $B$ cycle). For example, by inserting a flux quantum
$\frac{hc}{2e}$ through the cycle $A$, an electromotive force is
induced along it with a consequent transport of an electron along
the $B$ cycle.

The foundations of such an issue can be found in the celebrated
Laughlin gauge argument \cite{gauge} which runs as follows. Let us
focus on the geometry proposed by Laughlin, that is a ribbon of
two-dimensional system bent into a loop of circumference $L$ and
embedded everywhere by a strong magnetic field $\vec{B}$ normal to
its surface (see Fig. 3). Let us also put a small solenoid at the
center of the loop, as shown in the figure, and assume that an
energy gap separates the ground state from the excited states. In
order to force the system to produce Hall current let us also
assume that electrons can be fed in at one edge and taken away
from the other. Now we switch on the solenoid and adiabatically
increase the magnetic flux from zero to $\Phi_{0}=\frac{hc}{e}$.
Because of the energy gap, the system remains in a ground state
which may be different from the original one. If the ground state
is non degenerate, by gauge invariance the system simply returns
to the initial state. Because of the phase coherence of the wave
function of the system around the loop, the net result of such a
process will be the transfer of $N_{0}$ electrons from one edge to
the other. The energy increase due to this transfer is
\cite{gauge}
\begin{equation}\label{l1}
  \Delta U=N_{0}eV_{H}
\end{equation}%
where $V_{H}$ is the potential drop from one edge to another. The
Hall current is
\begin{equation}\label{l2}
  I_{H}=\frac{\partial U}{\partial \Phi}=\frac{\Delta
  U}{\Phi_{0}}=\frac{V_{H}N_{0}e^{2}}{h}
\end{equation}%
and the Hall conductance is
\begin{equation}\label{l3}
  \sigma_{H}=\frac{I_{H}}{V_{H}}=\frac{N_{0}e^{2}}{h}.
\end{equation}%
In this way quantization of the Hall conductance has been
reproduced for integer fillings and the argument has been
generalized also to fractional conductance \cite{gauge}. So in the
following we keep in mind this line of reasoning and then produce
a potential drop between the four terminals of our bilayer system
by adiabatic insertion of a magnetic flux quantum which results in
the transport of electrons on each edge and between edges. This
allows us to study transport properties.

\begin{figure}
  \includegraphics[height=.3\textheight]{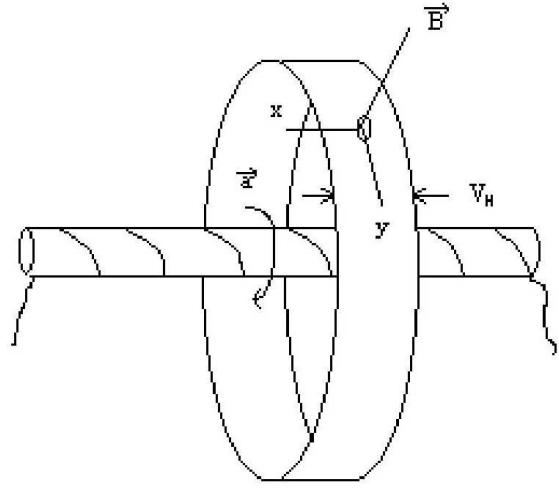}
  \caption{The loop and the solenoid in the geometry by Laughlin.}
\end{figure}

We focus in particular on the torus topology, where the transport
properties can be precisely described in terms of the action of
magnetic translations on the conformal blocks of the untwisted and
twisted sector respectively. Their explicit description can be
realized by standard calculations on the characters of the TM
given in refs \cite{cgm4}. In this letter we just recall that the
characters are given in terms of opportune Jacobi theta functions
with characteristics $\theta \left[
\begin{array}{c}
\lambda  \\
0%
\end{array}%
\right] \left( qw^{\left( i\right) }|2q\tau \right) $, where $\tau
$ is the modular parameter of the torus, $w^{\left( i\right)
}=x^{\left( i\right)
}+y^{\left( i\right) }\tau $ is the torus coordinate of the electron and $q=p+1$%
. Magnetic translations on the $i$-layer along the two cycles $A$
and $B$ are described by exponential of differential operators
acting on the $w$ dependence of the characters. In the bilayer
system the states belong to the $1/2$ representation of the
$su(2)$ pseudospin group. The TM on the torus keeps track of these
pseudospin configurations by the $w$ dependence of the
characters, whose charged and neutral components are described in terms of the layers variables $w^{(1)}$, $w^{(2)}$ as $%
w_{c}=(w^{(1)}+w^{(2)})/2$ and $w_{n}=(w^{(1)}-w^{(2)})/2$
respectively.

So the two configurations, given above, without tunneling are
described on the torus by the following translations on the
charged and neutral $w$ coordinate. In the non crossed case (Fig.
2a) the potential $V_{AC}$ ($V_{BD}
$) generates a translation along the first (second) layer, on the variable $%
w^{(1)}$ $(w^{(2)})$, and it results $\Delta w_{c}\propto
V_{AC}+V_{BD}$ and
$\Delta w_{n}\propto V_{AC}-V_{BD}$, while in the crossed case (Fig. 2b) $%
\Delta w_{c}\propto V_{AD}+V_{BC}$ and $\Delta w_{n}\propto
V_{AD}-V_{BC}$. At this point the study of the transport
properties follow by standard
analysis \cite{tran}. Let us point out that a purely neutral translation $%
w_{p}$ with $w^{(1)}=-w^{(2)}$ creates the topological defect (and
relates
the edges switching to the large unbalance phenomenon predicted in \cite%
{edge1}). In fact the twisted sector can be realized by a suitable
neutral translation starting from the untwisted one and its
explicit expression and derivation will be addressed in a
forthcoming publication \cite{tran}. Finally in the presence of
localized tunneling ($t_{p}\neq 0$)\ between the layers
hybridization takes place. In fact that experimentally corresponds
to an equilibration process between the two edge states and
results into a breaking of the symmetry of the balanced system
described by the TM, due to the breaking of pseudospin symmetry.
To take that into account the boundary CFT technology was used in
\cite{noi}, obtaining the characters of the system in the presence
of both tunneling and topological defects.

Let us now discuss the transport properties in these unbalanced
cases, by describing the tunneling as a small perturbation to the
TM, and focus our attention to the terminal $AD$ in the crossed
case. The working points are different for the untwisted and the
twisted configurations. In the first case the term in eq.
(\ref{tunnel}) for $t_{p}\ll 1$ is a weak perturbation of the
background characterized by $V_{AD}=V_{BC}=0$ while in the second
one it has $V_{AC}=V_{BD}=0$. The $I-V$ characteristics depends
strongly on that. We obtain a different conductance for the two
cases. In particular for TBC in the absence of an in plane
magnetic field the driving voltage $V_{AD}$ puts the bilayer edges
at different chemical potentials and then the ratio
of the $AD$ terminal current to $V_{AD}$ is equal to the Hall conductance $%
\sigma _{H}=\frac{e^{2}}{2h}$, of the single layer. Conversely,
when the two layers are coupled via the in plane magnetic field,
the tunneling of the charge carriers results into a loss in the
$AD$ terminal current. The net result is a negative contribution
to the current which adds to the previous term, producing a total
$AD$ terminal current, which for $p=0$, can be exactly evaluated
in a similar way as in \cite{fendley}, obtaining:
\begin{equation}
I_{AD}\left( V_{AD}\right)
=\frac{e^{2}V_{AD}}{2h}-\frac{eT_{B}}{h}\arctan
\frac{eV_{AD}}{2T_{B}},
\end{equation}%
where $T_{B}=C_{1}t_{P}^{1/\left( 1-\nu \right) }$ is the analogue
of the Kondo temperature, depending on the external parallel
magnetic field, $C_{1}$
is a non-universal constant and $\nu $ is the filling. In this case $\nu =%
\frac{1}{2}$, for the single layer, but the argument can be
generalized to a wide class of fillings.

\begin{figure}
  \includegraphics[height=.3\textheight]{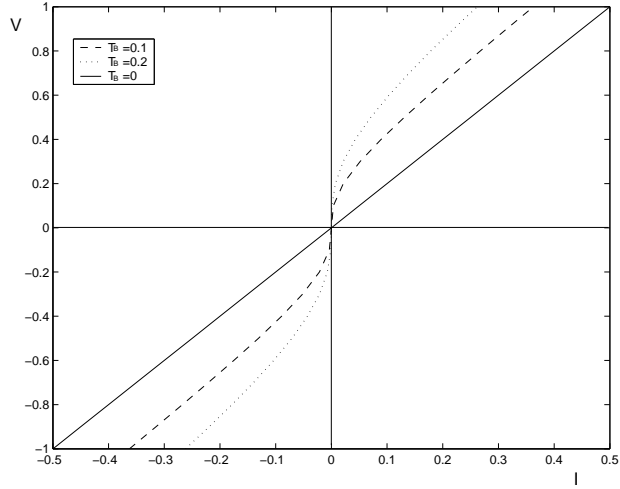}
  \caption{I-V
characteristics of the bilayer system in the twisted sector for
different values of $T_{B}$.}
\end{figure}

The non linear behavior of the tunneling characteristics follows
by standard analysis (ref. \cite{fendley}). Indeed for $T_{B}=0$
the characteristics has a linear behavior as for the transport in
a single layer. Moreover in plane magnetic field removes the twist
(topological defect) and re-establishes the non-linear structure
characterizing the tunneling phenomenon. Let us notice also that
our system is spinless (or fully polarized) while the experimental
results in \cite{deviatov1} are obtained for spin resolved
systems.
Therefore we reproduce only the negative branch of the curves given in \cite%
{deviatov1}. No gap is obtained for positive $V_{AD}$.

\section{Conclusions}

In conclusion we point out that the evidence of topological
defects, resulting from TBC, is theoretically indispensable for
the consistency of our CFT approach to the QHE. It is implied by
the $m$-reduction technique.

The presence of topological defects in a double layer induces flux
fractionalization described by the special $w_{p}$ translation and
is responsible for linear conduction between different edges with
a quantized value of the slope. In \cite{noi} the stability of the
different ground-states was studied by means of the boundary
entropy g. It was also related to dissipative quantum mechanics.
This is an interesting interpretation of our theoretical results
in connection also with the phase-transition between fully
polarized/unpolarized pseudospin vacua, in analogy to the observed
spin phase-transitions. Our theory for $p\neq 0$ predicts a
breaking of the composite fermion picture with a different
behavior for the fluxes (vortices), which are not sensible to the
topological defect.

We point out that the results of this letter are very general and
are relevant for different areas of condensed matter systems at
low dimensions. It has been shown that there is a close relation
between the existence of topological defects and flux
fractionalization in fully frustrated Josephson junction ladders
\cite{JJL}. Furthermore topological defects have been also
introduced in the description of dissipation in systems with impurities \cite%
{noi}.

\section*{Acknowledgements}

A. Naddeo kindly acknowledges the organizers of the X Training
Course in the Physics of Correlated Electron Systems, A. Avella,
F. Mancini and M. Marinaro, for giving her the possibility to
participate in the stimulating atmosphere of the school.

\end{document}